\documentclass[a4paper]{article}
\usepackage{INTERSPEECH2019}
\usepackage{graphicx}
\usepackage{amssymb, amsmath, bm, mathtools,array}
\usepackage{textcomp}
\usepackage{hyperref}
\usepackage{verbatim}
\usepackage{lipsum} 
\usepackage{multirow}
\usepackage{siunitx}
\usepackage{diagbox}
\newcommand{\bs}[1]{\boldsymbol{#1}}

\newcommand{\ceil}[1]{\left\lceil #1 \right\rceil}

\RequirePackage{color}\definecolor{BLUE}{rgb}{0,0,0} 

\usepackage[caption=false,font=footnotesize]{subfig}

\title{Neural Harmonic-plus-Noise Waveform Model with Trainable Maximum Voice Frequency for Text-to-Speech Synthesis}
\name{Xin Wang$^1$, Junichi Yamagishi$^{1,}$$^2$}
\address{
  $^1$National Institute of Informatics, Japan, $^2$CSTR, University of Edinburgh, UK}
\email{wangxin@nii.ac.jp, jyamagis@nii.ac.jp}

\begin{document}

\maketitle
\begin{abstract}
Neural source-filter (NSF) models are deep neural networks that produce waveforms given input acoustic features.
They use dilated-convolution-based neural filter modules to filter sine-based excitation for waveform generation, 
which is different from WaveNet and flow-based models. 
One of the NSF models, called harmonic-plus-noise NSF (h-NSF) model, uses separate pairs of source and neural filters to generate  harmonic and noise waveform components. It is close to WaveNet in terms of speech quality while being superior in generation speed. 

The h-NSF model can be improved even further. 
While h-NSF merges the harmonic and noise components using pre-defined digital low- and high-pass filters, it is well known that the maximum voice frequency (MVF) that separates the periodic and aperiodic spectral bands are time-variant. 
Therefore, we propose a new h-NSF model with time-variant and trainable MVF. 
We parameterize the digital low- and high-pass filters as windowed-sinc filters and predict their cut-off frequency (i.e., MVF) from the input acoustic features. 
Our experiments demonstrated that the new model can predict a good trajectory of the MVF and produce high-quality speech for a text-to-speech synthesis system.

\end{abstract}

\section{Introduction}
In text-to-speech (TTS) systems using statistical parametric speech synthesis \cite{ref:Zen09}, 
neural-network (NN)-based models have been introduced to both the front-end text analyzer and the back-end acoustic models
\cite{yao2015sequence,zen2013statistical,kaneko2017generative,henter2016robust,wang2018autoregressive}.
The recent trend is to replace the signal-processing-based vocoder with a neural waveform model, a component that generates a waveform from the acoustic features predicted by the acoustic models. 

A well-known neural waveform model called WaveNet-vocoder \cite{Tamamori2017} uses a dilated convolution (CONV) network \cite{oord2016wavenet} to produce the waveform samples in an autoregressive (AR) manner, i.e., generating the current waveform sample with the previously generated samples as condition. 
Although WaveNet outperformed traditional vocoders \cite{wangICASSP2018}, its sequential generation process is prohibitively slow. 
Flow-based models \cite{prenger2018waveglow,pmlr-v80-oord18a,ping2018clarinet} convert a noise sequence into a waveform in one shot. 
However, some of them require sequential processing during training \cite{prenger2018waveglow}, which dramatically increases the training time \cite{okamoto-waveglow}. Others use  knowledge distilling to transfer the knowledge from an AR WaveNet to a flow-based student model, which is complicated in implementation. 

We recently proposed neural source-filter (NSF) waveform models which require neither AR structure, knowledge distilling, nor flow-based methods \cite{wang2018neural}.
The NSF models generally use three modules to generate a waveform: 
a conditional module that upsamples input acoustic features such as F0 and Mel-spectrograms,
a source module that outputs a sine-based excitation given the F0, 
and a filter module that uses dilated-CONV blocks to morph the excitation into a waveform.
The models are trained to minimize the spectral amplitude distance between the generated and natural waveforms. 
Without the flow-based approach, the NSF models are easy to implement and train.
Without the AR structure, the NSF models are at least 100 times faster than WaveNet for waveform generation \cite{nsf-all}. 

An NSF model called harmonic-plus-noise NSF (h-NSF) inherits the efficiency of the NSF models and demonstrates comparable or better performance than WaveNet and other NSF models on a Japanese dataset \cite{nsf-all}.  The core idea of h-NSF is to use separate pairs of the source and neural filter modules to generate harmonic and noise waveform components before merging the two components into an output waveform by using pre-defined finite impulse response (FIR) filters.  The harmonic-plus-noise architecture of h-NSF improves the quality of the generated waveforms, especially on unvoiced sounds.

It is well known that the speech spectrum can be roughly divided into periodic and aperiodic bands by a maximum voice frequency (MVF) \cite{stylianou1996harmonic}. 
Although MVF is time-variant, our h-NSF chooses one of the two pre-defined MVF values (i.e., the cut-off frequency of FIR filters) according to the voicing status of the sound. 
In this paper, we propose a new h-NSF model with trainable MVF.
This new model parameterizes the FIR filters as windowed-sinc filters \cite{Smith:1997:SEG:281875} and predicts their MVF values from the input acoustic features. Our experiments demonstrated that the new h-NSF can predict the MVF reasonably well on the basis of the voicing status. The quality of the generated waveforms has improved without any detriment to the generation speed.

Because the new h-NSF model replies on windowed-sinc filters, we refer to it as \textit{sinc-h-NSF}, while the previous h-NSF is referred to as \textit{base-h-NSF}. In Section~\ref{seq:review}, we explain the details of base-h-NSF. In Section~\ref{seq:proposed_h_nsf}, we describe sinc-h-NSF. In Section~\ref{seq:exp}, we compare the two h-NSF models with WaveNet in experiments. In Section~\ref{sec:conclude}, we draw a conclusion.

\section{Review of base-h-NSF model}
\label{seq:review}
A neural waveform model converts input acoustic features into an output waveform.
Let us denote the input acoustic feature sequence as $\bs{c}_{1:B}=\{\bs{c}_{1}, \cdots  \bs{c}_B\}$, where 
$\bs{c}_b\in\mathbb{R}^{D}$ is the feature vector for the $b$-th frame.  
We then use $\bs{o}_{1:T}=\{o_1, \cdots, o_T\}$ and $\widehat{\bs{o}}_{1:T}$ to denote the natural and generated waveforms, respectively. 
Here, $T$ is the waveform length and $o_t\in\mathbb{R}$ is the waveform value at the $t$-th sampling point. 

\begin{figure}[!t]
\centering
{\includegraphics[width=\columnwidth]{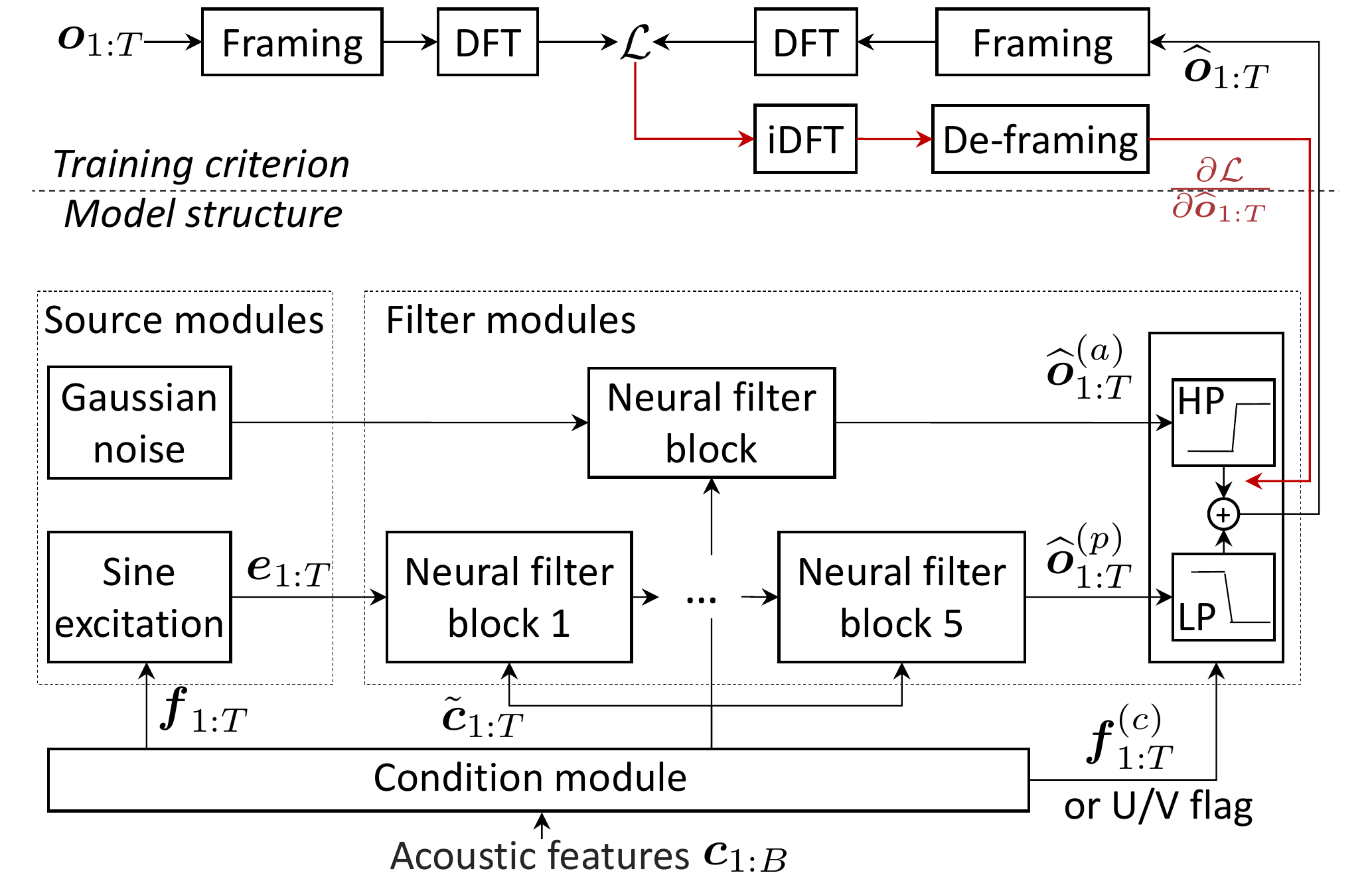}
}
\vspace{-6mm}
\caption{General network structure for baseline and proposed trainable h-NSF models. LP and HP denote low- and high-pass FIR filters, respectively. Red arrows denote gradients.}
\label{fig:system_overall}
\end{figure}

In our previous work, we proposed NSF models \cite{wang2018neural} to convert $\bs{c}_{1:B}$ into $\widehat{\bs{o}}_{1:T}$.
The NSF models use three types of modules:
a source to produce an excitation signal, a neural filter to convert the excitation into $\widehat{\bs{o}}_{1:T}$, and a condition part to processes input $\bs{c}_{1:B}$ for the other two modules. The training is conducted by minimizing the spectral distance between $\widehat{\bs{o}}_{1:T}$ and ${\bs{o}}_{1:T}$ \cite{wang2018neural}.
Base-h-NSF model introduces harmonic-plus-noise architecture to the NSF framework (Figure~\ref{fig:system_overall}).
The details of base-h-NSF are explained in the following sections.
%

\subsection{Condition module}
\label{sec:model_cond}
The condition module is the bedrock of base-h-NSF. Its basic task is to upsample the frame-rate acoustic features to the waveform rate.
As Figure~\ref{fig:system_overall} shows, the condition module processes three types of features\footnote{
There is one alternative feature $\bs{f}_{1:T}^{(c)}$ that is used by the trainable h-NSF model proposed in this paper (Section~\ref{seq:proposed_h_nsf}). It is not used by the baseline h-NSF models.}:
the upsampled F0 sequence $\bs{f}_{1:T}$ for the source module, 
the upsampled and transformed acoustic feature sequence $\widehat{\bs{c}}_{1:T}$ for the neural filter module, and the upsampled unvoiced/voiced (U/V) flag for the FIR filters. 

Suppose each frame of the input $\bs{c}_{1:B}$ contains an F0 datum $f_b\in{\mathbb{R}}_{\geq{0}}$ and a Mel-spectrum $\bs{s}_b$, i.e., $\bs{c}_{b}=[f_b, \bs{s}_b^{\top}]^{\top}$.
Then, it is straightforward to upsample the F0 sequence $\{f_1, \cdots, f_B\}$ of length $B$ into $\bs{f}_{1:T}$ of length $T$ by simply copying each $f_b$ for $\ceil{T/B}$ times. 
Similarly, the U/V flag sequence can be upsampled after determining the U/V from the $f_b$ (e.g., voiced if $f_b>0$ or unvoiced if $f_b=0$).
For $\widehat{\bs{c}}_{1:T}$, the condition module first transforms the sequence of $\bs{s}_b$ using two hidden layers: a bi-directional long short-term memory (Bi-LSTM) layer with a layer size of 64 and a 1-D convolution (CONV) layer with a layer size of 63 and a window size of 3. 
After that, it concatenates the output feature vector with the F0 and upsamples it as $\tilde{\bs{c}}_{1:T}$, where $\tilde{\bs{c}}_{t}\in\mathbb{R}^{64}, \forall{t}\in\{1,\cdots, T\}$. 

\begin{figure}[!t]
\centering
\subfloat{\includegraphics[width=\columnwidth]{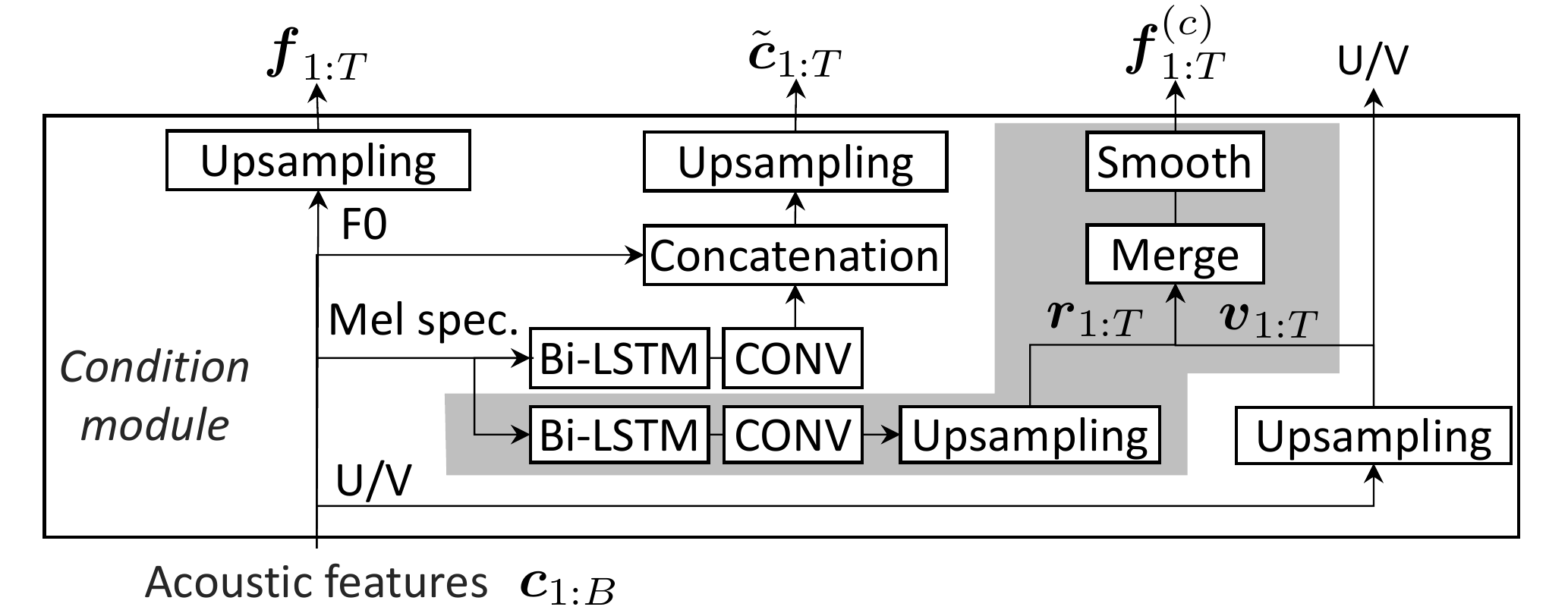}
}
\hfil
\subfloat{\includegraphics[width=\columnwidth]{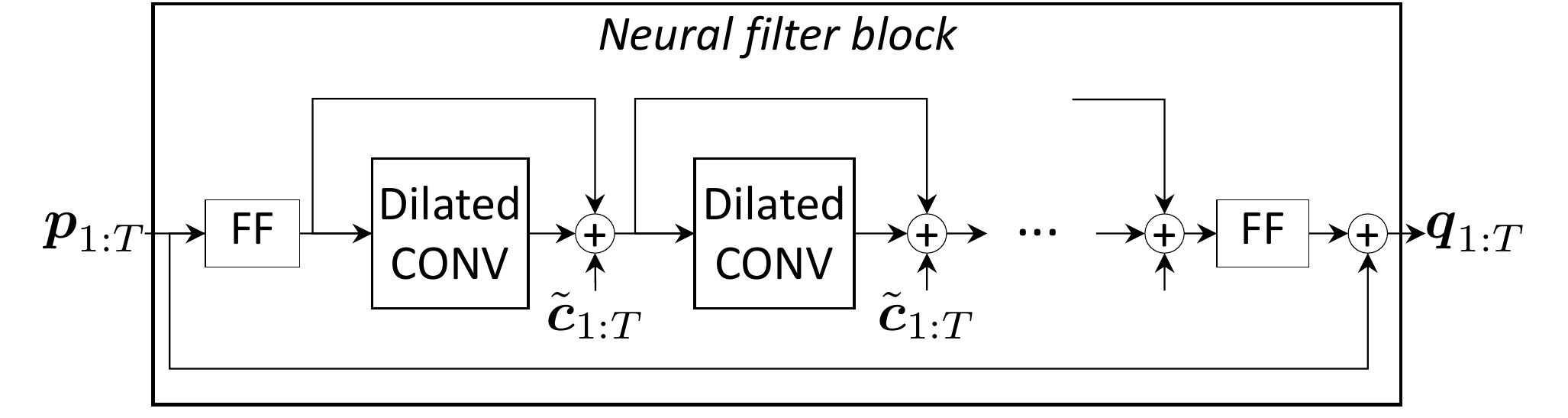}
}
\vspace{-3mm}
\caption{Condition module (top) and neural filter block (bottom).
{FF}, {Bi-LSTM}, and {CONV} denote feedforward, bi-directional LSTM, and convolutional layers, respectively. Layers in shaded area are only used to compute $\bs{f}_{1:T}^{(c)}$ for the proposed new h-NSF model.}
\label{fig:cond_filter}
\end{figure}

\subsection{Source modules}
The base-h-NSF model contains two source modules.
One module generates Gaussian noise excitation for the noise waveform component, while the other generates a sine-based excitation signal $\bs{e}_{1:T}$ for the harmonic component.

We briefly explain the sine-based excitation.
Given the upsampled F0 sequence $\bs{f}_{1:T}$, a sine waveform that carries the F0 or the $i$-th harmonic can be generated as
\begin{align}
{e}_t^{<i>} = \begin{dcases}
\alpha\sin(\sum_{k=1}^{t}2\pi\frac{i{f}_k}{N_s} + {\phi}) + {n}_t, &\text{if }{f_t}>0 \\ 
\frac{\alpha}{3\sigma} {n}_t, & \text{if } f_t = 0\\ 
\end{dcases},
\label{eq:sine}
\end{align}
where $\phi\in[-\pi, \pi]$ is a random initial phase, $N_s$ is a waveform sampling rate, and ${n}_t \sim \mathcal{N}(0, \sigma^2)$ is Gaussian noise. 
Note that ${e}_t^{<i>}$ is a Gaussian noise in unvoiced regions where $f_t=0$.
The hyper-parameter $\alpha$ adjusts the amplitude of ${e}_t^{<i>}$, while $\sigma$ is the standard deviation of the Gaussian distribution.
We set $\sigma=0.003$ and $\alpha=0.1$ \cite{nsf-all}. 

We set $I=8$ for base-h-NSF, i.e., fundamental tone and seven higher harmonics.
A feedforward (FF) layer is then used to merge them into excitation ${\bs{e}_{1:T}}=\text{tanh}(\sum_{i=1}^{I}w_i{\bs{e}}_{1:T}^{<i>} + w_b)$, where ${e}_t\in\mathbb{R}, \forall{t}\in\{1,\cdots, T\}$. Note that $\{w_1, \cdots w_I, w_b\}$ are the FF layer's weights.

\subsection{Filter modules}
The filter modules of the base-h-NSF can be described in three parts.
The first part uses one neural filter block to convert Gaussian noise into a noise waveform component $\widehat{\bs{o}}_{1:T}^{(a)}$, while the second part uses five blocks to convert $\bs{e}_{1:T}$ into a harmonic waveform component $\widehat{\bs{o}}_{1:T}^{(p)}$. The third part uses FIR filters to merge $\widehat{\bs{o}}_{1:T}^{(a)}$ and $\widehat{\bs{o}}_{1:T}^{(p)}$ into the output waveform $\widehat{\bs{o}}_{1:T}$.

The neural filter block is plotted in Figure~\ref{fig:cond_filter}. 
Suppose the input signal is $\bs{p}_{1:T}$, where ${p}_t\in\mathbb{R}, \forall{t}\in\{1,\cdots,T\}$\footnote{For the 1st block that receives $\bs{e}_{1:T}$ as input,   $\bs{p}_{1:T} = \bs{e}_{1:T}$.}. 
Each $\bs{p}_{t}$ is first expanded to 64 dimensions through an FF layer, then processed by a dilated-CONV layer with 64 output channels, and finally summed with the output of the dilated-CONV layer and the conditional feature $\tilde{\bs{c}}_{1:T}$.
This process is repeated 10 times; the final output sequence is transformed back into a 1-dimensional signal through a FF layer and then summed with $\bs{p}_{1:T}$. Note that the dilation size of the $k$-th dilated-CONV layer is $2^{\text{mod(k-1, 10)}}$, and its filter size is set to 3. 


After the neural filter blocks generate $\widehat{\bs{o}}_{1:T}^{(p)}$ and $\widehat{\bs{o}}_{1:T}^{(a)}$, the base-h-NSF uses low- and high-pass FIR filters to mix them as the output waveform $\widehat{\bs{o}}_{1:T} = \text{Low-pass}(\widehat{\bs{o}}_{1:T}^{(p)}) + \text{High-pass}(\widehat{\bs{o}}_{1:T}^{(a)})$.
In implementation, 
we switch the cut-off frequency (-3 dB) of the FIR filters on the basis of the U/V flag.
In voiced regions, the cut-off frequency values for the low- and high-pass filters are 5 kHz and 7 kHz, respectively.
In unvoiced regions, they are 1 kHz and 3kHz. The filter coefficients are calculated in advance \cite{parks1972chebyshev} and fixed in the model.

\section{Proposed h-NSF model with trainable maximum voice frequency}
\label{seq:proposed_h_nsf}
The cut-off frequency of the FIR filters in base-h-NSF is manually specified and only changes according to 
the voicing conditions.
In classical harmonic-plus-noise models, however, the cut-off frequency is assumed to be time-variant  \cite{stylianou1996harmonic, drugman2014maximum}. 
It is thus reasonable to try time-variant FIR filters with a cut-off frequency predicted from the input acoustic features.

\begin{figure}[t]
\centering
\includegraphics[width=\columnwidth]{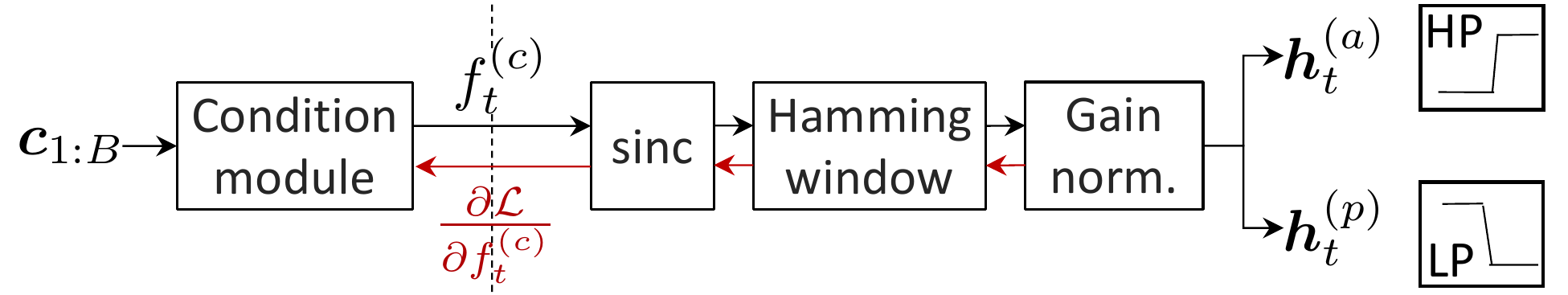}
\vspace{-5mm}
\caption{Procedure to derive low-pass (LP) and high-pass (HP) filter coefficients in proposed trainable h-NSF model.}
\label{fig:forward}
\end{figure}

The proposed h-NSF model is identical to the base-h-NSF except for the procedure to calculate the time-variant cut-off frequency for the  FIR filters.
Suppose we are using filters of order $M$, and their coefficients at time $t$ are $\bs{h}_{t}^{(p)} = \{h_{t,0}^{(p)}, \cdots, h_{t,M}^{(p)}\}$ and $\bs{h}_{t}^{(a)} = \{h_{t,0}^{(a)}, \cdots, h_{t,M}^{(a)}\}$, respectively.  
Given the periodic  and aperiodic components $\{\widehat{\bs{o}}_{1:T}^{(p)}, \widehat{\bs{o}}_{1:T}^{(a)}\}$, the output waveform $\widehat{{o}}_t$ at the $t$-th time step can be merged as
\begin{equation}
\widehat{{o}}_t = \sum_{m=0}^{M-1}\widehat{{o}}_{t-m}^{(p)}h_{t,m}^{(p)} + \sum_{m=0}^{M-1}\widehat{{o}}_{t-m}^{(a)}h_{t,m}^{(a)}.
\label{eq:output_waveform}
\end{equation}
Our goal is to predict $\{\bs{h}_{t}^{(p)},\bs{h}_{t}^{(a)}\}$ from the acoustic features $\bs{c}_{1:B}$.
For this purpose, we use a two-step procedure as Figure~\ref{fig:forward} plots.
First, the condition module predicts normalized cut-off frequency $f_t^{(c)}\in(0,1)$\footnote{Being normalized means that $f_t^{(c)}$ is equal to physical cut-off frequency (Hz) divided by Nyquist frequency.} given $\bs{c}_{1:B}$. After that, $\{\bs{h}_{t}^{(p)},\bs{h}_{t}^{(a)}\}$ are calculated from $f_t^{(c)}$. 
During back propagation, the gradients are computed and propagated backwards. 

\subsection{Forward computation}
\subsubsection{Predicting cut-off frequency}
\label{seq:cut_off_f}
Because the MVF of a sound is influenced by its voicing status, we take the U/V flag into consideration and revise the condition module of the base-h-NSF in order to predict $f_t^{(c)}$ from $\bs{c}_{1:B}$.
As the shaded area of Figure~\ref{fig:cond_filter} shows, a Bi-LSTM layer and a CONV layer with a tanh activation function are added to predict a signal that will be upsampled to $\bs{r}_{1:T}$, where $r_t\in{(-1, 1)}, \forall{t}\in\{1,\cdots,T\}$. 
Meanwhile, the U/V flag is upsampled as signal $\bs{v}_{1:T}$. 
We then set $v_t=0.7$ if the $t$-th time step is voiced or $v_t=0.3$ for an unvoiced time step\footnote{These values are references suggesting that, for example, the MVF of voiced sounds is around 5.6 kHz (=0.7 * 8 kHz). We can scale or shift these values when we merge $v_t$ with $r_t$.}.


With ${v}_{t}\in\{0.7, 0.3\}$ and ${r}_{t}\in(-1, 1)$, we can fuse them into the output $f_t^{(c)}\in(0, 1)$ in various ways.
Without loss of generality, we design a fusion function as
\begin{equation}
f_t^{(c)} = \mathcal{F}(a{v}_{t} + b{r}_{t} + c),
\label{eq:merge_v_r}
\end{equation}
where $\{a,b,c\}$ can be trainable parameters or fixed hyper-parameters.
$\mathcal{F}(\cdot)$ can be a sigmoid function or an identity function $f(x)=x$ if $a{v}_{t} + b{r}_{t} + c$ is already between 0 and 1.

How we define Equation~(\ref{eq:merge_v_r}) depends on our prior knowledge about the MVF and its relationship with the voicing status. For example, we may use three definitions listed in Table~\ref{tab:merge}.
The first definition ensures that $f_t^{(c)}\in(0.1,0.5)$ in voiced time steps while $f_t^{(c)}\in(0.5,0.9)$ in unvoiced time steps.
The second definition only replies on the $r_t$, i.e., predicting the MVF from conditional features without any prior knowledge of the voicing status. 
Finally, the last definition learns the weight to combine the $v_t$ and $r_t$. 
We compare these three definitions in our experiments.
Note that to prevent the abrupt change of the filter frequency response, $\bs{f}_{1:T}^{(c)}$ is smoothed by taking the time domain average over a window size of 5 ms.

\begin{table}[!t]
\vspace{-2mm}
\caption{Four possible definitions of $f_t^{(c)} = \mathcal{F}(a{v}_{t} + b{r}_{t} + c)$ (Equation~(\ref{eq:merge_v_r})) to merge ${v}_{t}\in\{0.7, 0.3\}$ and ${r}_{t}\in(-1, 1)$.}
\vspace{-3mm}
\begin{center}
\begin{tabular}{lcccc}
\hline\hline
Definition &  $\mathcal{F}(x)$ & $a$ & $b$  & $c$ \\
\hline
$f_t^{(c)}=v_t + 0.2r_t$ & $x$ & 1 & 0.2 & 0 \\
$f_t^{(c)}=0.5r_t+0.5$ & $x$ & 0 & 0.5 & 0.5 \\
$f_t^{(c)}= \mathcal{F}(a{v}_{t} + b{r}_{t} + c)$ & sigmoid & \multicolumn{3}{c}{trainable} \\
\hline
\hline
\end{tabular}
\end{center}
\label{tab:merge}
\end{table}

\subsubsection{Windowed-sinc filters}
Given $f_{t}^{(c)}$, we follow the standard procedure to design 
windowed-sinc filters and calculate $\{\bs{h}_{t}^{(p)}, \bs{h}_{t}^{(a)}\}$.
Suppose the filter length $M$ is an odd number, and the index of the filter coefficient is centered around 0, i.e., 
$n\in\{-\frac{M-1}{2}, -\frac{M-1}{2} + 1, \cdots, 0, \cdots, \frac{M-1}{2}\}$.
Given the $f_{t}^{(c)}$ at the $t$-th time step, 
a coefficient can be computed for each $n$ based on the sinc function and the Hamming window $\text{Hamm}(\cdot)$
\footnote{We used the Hamming window as SincNet did \cite{ravanelli2018speaker}. However, due to historical reasons, we used Hann window when calculating the spectral distances (i.e., Framing in Figure~\ref{fig:system_overall}.) }:
\begin{equation}
\begin{split}
\tilde{h}^{(p)}_{t,n} & = f_{t}^{(c)}\text{sinc}(\pi{f_{t}^{(c)}}n)\text{Hamm(n)} \\
& = \frac{\sin(\pi{f_{t}^{(c)}}n)}{\pi{n}} \big(0.54 + 0.46\cos(\frac{2\pi{n}}{M})\big).
\label{eq:low_pass_1}
\end{split}
\end{equation}
The desired filter coefficient $\bs{h}_{t}^{(p)}$ can then be calculated after gain normalization and index shifting from $n$ to $m$:
\begin{equation}
{h}^{(p)}_{t,m} = \frac{\tilde{h}^{(p)}_{t,m-\frac{M-1}{2}}}{\sum_{n={-\frac{M-1}{2}}}^{\frac{M-1}{2}}\tilde{h}^{(p)}_{t,n}}
\label{eq:low_pass_2}
\end{equation}
Note that gain normalization makes the gain of the low-pass filter equal to 1 at 0 Hz. The index shift from $n\in\{-\frac{M-1}{2}, \cdots, 0, \cdots, \frac{M-1}{2}\}$ to $m\in\{0, \cdots, {M}\}$ makes the filter causal.
Similarly, the high-pass filter coefficients are deterministically computed by:
\begin{equation}
\tilde{h}^{(a)}_{t,n} = \Big(\frac{\sin(\pi{n})}{\pi{n}} - \frac{\sin(\pi{f_{t}^{(c)}}n)}{\pi{n}} \Big) \text{Hamm}(n),
\end{equation}
\begin{equation}
{h}^{(a)}_{t,m} = \frac{\tilde{h}^{(a)}_{t,m-\frac{M-1}{2}}}{\sum_{n={-\frac{M-1}{2}}}^{\frac{M-1}{2}}\tilde{h}^{(a)}_{t,n}(-1)^{n}}.
\end{equation}

Trainable sinc-based FIR filters have been used in SincNet \cite{ravanelli2018speaker}.
While SincNet uses multiple time-invariant band-pass filters, we used time-variant low- and high-pass ones.
The cut-off frequency in SincNet is assumed to be the parameter of the network, but our network predicts it from conditional features. 

\subsection{Back propagation}
We need to calculate the gradients of $f_t^{(c)}$ w.r.t the loss function $\mathcal{L}$. 
By using the chain rule on Equation~(\ref{eq:output_waveform}), we first get
\begin{equation}
\frac{\partial{\mathcal{L}}}{\partial{f_t^{(c)}}} = \frac{\partial{\mathcal{L}}}{\partial{\widehat{o}_t}}\frac{\partial{\widehat{o}_t}}{\partial{f_t^{(c)}}} = \frac{\partial{\mathcal{L}}}{\partial{\widehat{o}_t}}\sum_{m=0}^{M-1}(\widehat{{o}}_{t-m}^{(p)} \frac{\partial{h_{t,m}^{(p)}}}{\partial{{f_t^{(c)}}}}  + \widehat{{o}}_{t-m}^{(a)} \frac{\partial{h_{t,m}^{(a)}}}{\partial{{f_t^{(c)}}}}).
\label{eq:bp_1}
\end{equation}
Then, based on Equations~(\ref{eq:low_pass_1}) and (\ref{eq:low_pass_2}), it can be shown that
\begin{equation}
\frac{\partial{h_{t,m}^{(p)}}}{\partial{{f_t^{(c)}}}} = \sum_{n}\frac{\partial{h_{t,m}^{(p)}}}{\partial{\tilde{h}_{t,n}^{(p)}}} \frac{\partial{\tilde{h}_{t,n}^{(p)}}}{\partial{{f_t^{(c)}}}} = \frac{\alpha_{t,m-\frac{M-1}{2}} - {h}^{(p)}_{t,m}\gamma_{t}^{(p)}}{\beta_{t}^{(p)}},
\label{eq:bp_2}
\end{equation}
where $\alpha_{t,n} = \text{Hamm}(n)\cos(\pi{f_{t}^{(c)}}n)$, $\beta_{t}^{(p)} = \sum_{n}\tilde{h}^{(p)}_{t,n}$, and 
$\gamma_{t}^{(p)} = \sum_{n}\alpha_{t,n}$.
Similarly, ${\partial{h_{t,m}^{(a)}}}/{\partial{{f_t^{(c)}}}}$ can be calculated as
\begin{equation}
\frac{\partial{h_{t,m}^{(a)}}}{\partial{{f_t^{(c)}}}} = \frac{{h}^{(a)}_{t,m}\gamma_{t}^{(a)} - \alpha_{t,m-\frac{M-1}{2}}}{\beta_{t}^{(a)}},
\label{eq:bp_3}
\end{equation}
where $\beta_{t}^{(a)} = \sum_{n}(-1)^{(n)}\tilde{h}^{(a)}_{t,n}$ and $\gamma_{t}^{(a)}=\sum_{n}(-1)^{n}\alpha_{t,n}$.

On the basis of Equations~(\ref{eq:bp_2}) and (\ref{eq:bp_3}), ${\partial{\mathcal{L}}}/{\partial{f_t^{(c)}}}$ in Equation~(\ref{eq:bp_1}) can be computed and propagated backwards. 

\section{Experiments}
\label{seq:exp}
\subsection{Data and feature configuration}
\label{sec:data}
For the experiment, we used the same data corpus and feature configuration as our previous work \cite{nsf-all}. 
Specifically, the corpus is a neural-style reading speech dataset from a Japanese female speaker.
The original speech waveforms were down sampled from 48 kHz to 16 kHz for the experiments. 

To train the neural waveform models, we randomly selected 9,000 utterances (15 hours) as the training set.
We then prepared a validation set with 500 randomly selected utterances and a test set with another 480 utterances. 
The acoustic features included the Mel-spectrograms of 80 dimensions and the F0 extracted using an ensemble of pitch estimators \cite{juvelaniibc}.
The frame shift of the acoustic features was 5 ms (200 Hz). 

Because we planned to evaluate the neural waveform models not only in copy-synthesis but also in TTS scenarios, 
we also extracted linguistic features from the transcripts to train acoustic models that predict the Mel-spectrogram and F0 from the text.
The linguistic features contained quin-phone identity, phrase accent type, etc. \cite{luong2018investigating}.
These features were then aligned against the acoustic feature sequences.

\subsection{Experimental models}
\label{sec:models}
We compared the following models in the experiment\footnote{Codes, scripts, and samples:\url{https://nii-yamagishilab.github.io/samples-nsf/nsf-v3.html}}:
\begin{itemize}
\item \texttt{WaveNet}: an AR WaveNet;
\item \texttt{base-h-NSF}: base-h-NSF using the fixed coefficients for the low- and high-pass filters;
\item \texttt{sinc1-h-NSF}: h-NSF with windowed-sinc filters and cut-off frequency $f_t^{(c)}=v_t + 0.2r_t$;
\item \texttt{sinc2-h-NSF}: h-NSF with windowed-sinc filters and cut-off frequency $f_t^{(c)}=0.5r_t+0.5$;
\item \texttt{sinc3-h-NSF}: h-NSF with windowed-sinc filters and cut-off frequency $f_t^{(c)}= \text{Sigmoid}(a{v}_{t} + b{r}_{t} + c)$;
\end{itemize}

\texttt{base-h-NSF} was trained in our previous work \cite{nsf-all}.
It used five dilated-CONV filter blocks (Figure~\ref{fig:cond_filter}) to generate the harmonic waveform component, and each block contained ten dilated-CONV layers. The $k$-th dilated-CONV layer had a dilation size of $2^{k-1}$. 
For the noise component, \texttt{base-h-NSF} used only one block. 
The three \texttt{sinc*-h-NSF} models used the same network structure as \texttt{base-h-NSF} except for the hidden layers to predict cut-off frequency for the time-variant FIR filters. The FIR filters used $M=31$.
All NSF models were trained using the sum of three spectral amplitude distances with framing and windowing configurations listed in Table~\ref{tab:dft_config}.

\begin{table}[!t]
\caption{Short-time analysis configurations for the spectral amplitude distance of NSF models}
\vspace{-6mm}
\begin{center}
\begin{tabular}{rccc}
\hline\hline
 & $\mathcal{L}_{1}$ & $\mathcal{L}_{2}$ & $\mathcal{L}_{3}$ \\
 \hline
DFT bins         &  512 & 128 & 2048 \\
Frame length  & 320 (20 ms) & 80 (5 ms)  & 1920 (120 ms) \\
Frame shift         & 80 (5 ms)   & 40 (2.5 ms) & 640 (40 ms) \\
\hline\hline
\multicolumn{4}{c}{Note: all configurations use the Hann window. } \\
\end{tabular}
\end{center}
\label{tab:dft_config}
\vspace{-5mm}
\end{table}

\begin{figure}[!t]
\centering
{\includegraphics[width=\columnwidth]{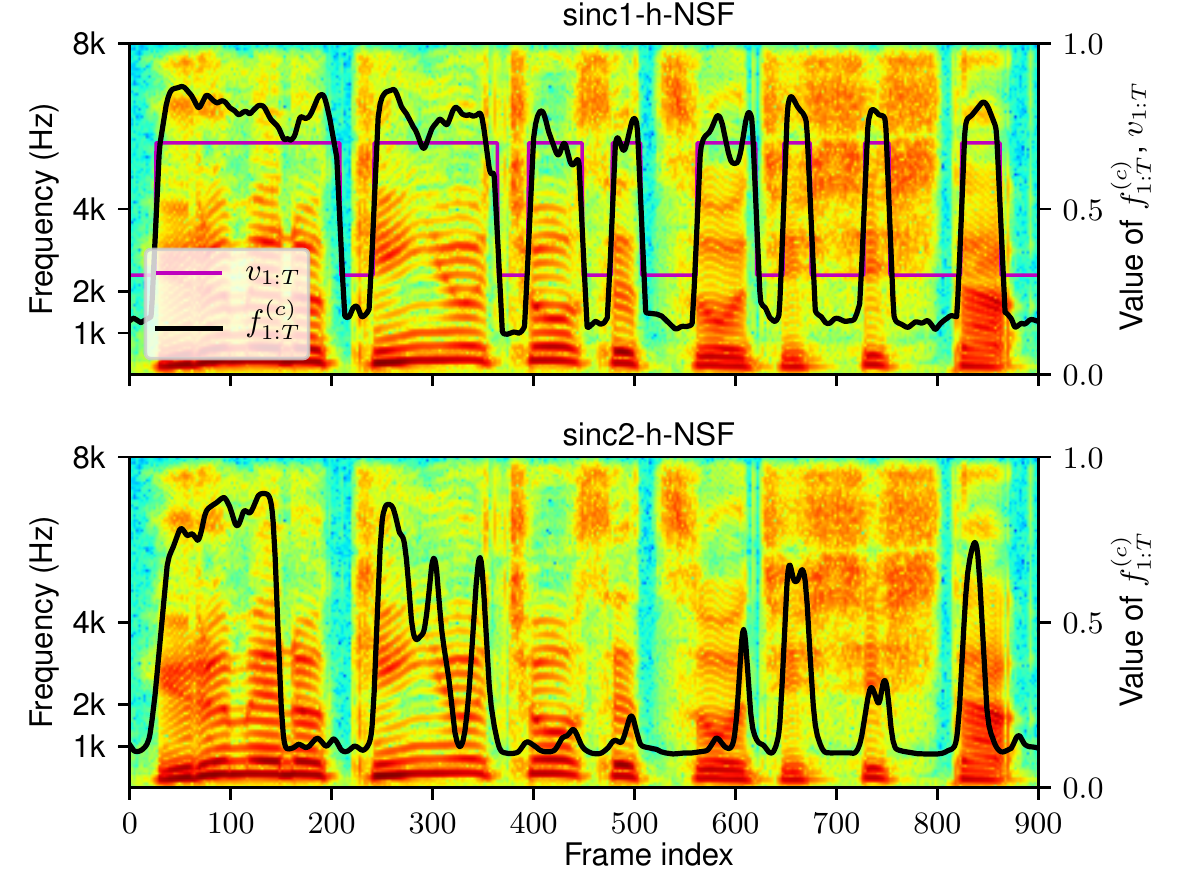}
}
\vspace{-8mm}
\caption{Predicted $\bs{f}_{1:T}^{(c)}$ and $\bs{v}_{1:T}$ when model was conditioned on natural acoustic features. Background is the spectrogram of natural waveform. Figure for \texttt{sinc3-h-NSF} was not plotted because the generated $\bs{f}_{1:T}^{(c)}$ was 1.0.}
\label{fig:mvf_ana}
\end{figure}

\begin{figure*}[!t]
\centering
{\includegraphics[width=\textwidth]{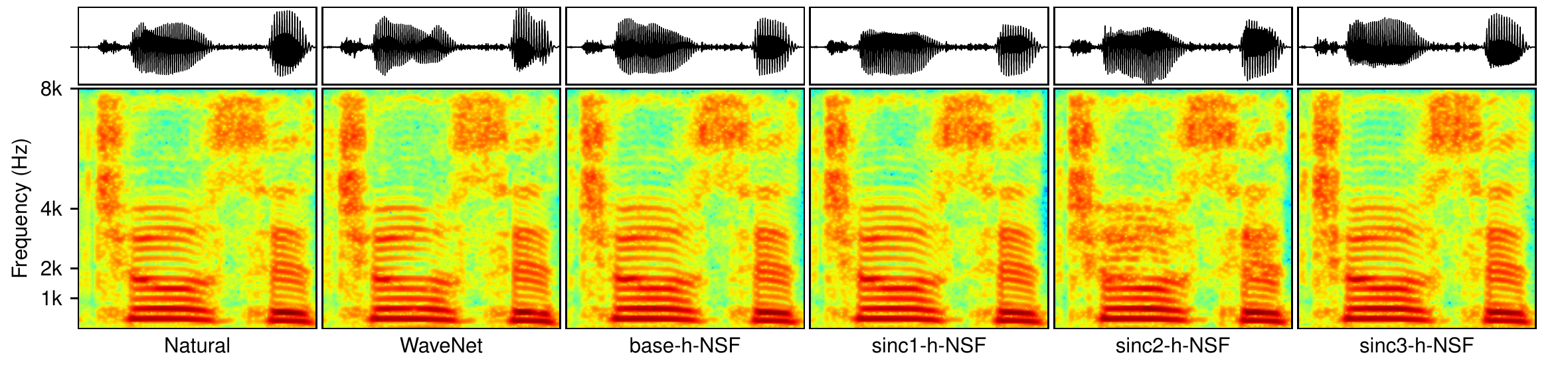}
}
\vspace{-8mm}
\caption{Generated waveforms given natural Mel-spectrogram and F0 for the same utterance as in Figure~\ref{fig:mvf_ana}. Only the 370th to 510th frames are plotted. }
\vspace{-5mm}
\label{fig:spec}
\end{figure*}

\texttt{WaveNet} was trained in our previous work \cite{wangICASSP2018}. 
It contained 40 dilated CONV layers, where the $k$-layer had a dilation size of $2^{\text{mod}(k-1, 10)}$.
\texttt{WaveNet} took both Mel-spectrogram and F0 as conditional features and generated 10-bit $\mu$-law quantized waveform values. 

To predict the acoustic features from the linguistic features, we used a deep neural AR F0 model \cite{wang2018autoregressive} for predicting the F0 and another deep AR model for the Mel-spectrogram. The acoustic feature sequences were generated given the duration aligned on the test set waveforms.

\subsection{Results and analysis}
\label{sec:results}
We first compared the predicted MVF from the \texttt{sinc*-h-NSF} models.
Figure~\ref{fig:mvf_ana} plots the predicted MVF trajectory and the natural waveform spectrogram.
Without using U/V, \texttt{sinc2-h-NSF} failed to predict MVF for some voiced regions, for example, from the 400-th to 500-th frames.
Although \texttt{sinc3-h-NSF} used the U/V, the function $\text{Sigmoid}(a{v}_{t} + b{r}_{t} + c)$ was saturated and produced 1.0 for all  time steps.
It seemed to be difficult to learn a trainable function to merge the u/v and the other acoustic features for MVF prediction.
MVF predicted from \texttt{sinc1-h-NSF} is in general consistent with the spectrogram, i.e., high MVF in voiced regions and low MVF in unvoiced regions. 
These results suggest that MVF can be predicted reasonably well by summing the U/V with a residual signal predicted from the input acoustic features.

\begin{figure}[!t]
\centering
{\includegraphics[width=\columnwidth]{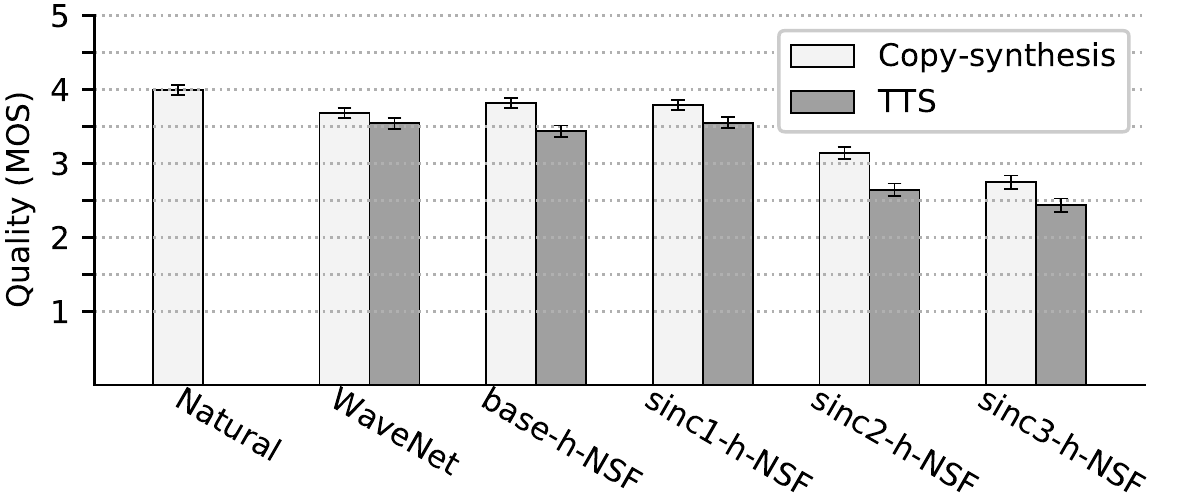}
}
\vspace{-6mm}
\caption{MOS scores of experimental models. Error bars at confidence level of 95\% are plotted.}
\label{fig:mos}
\end{figure}

We then compared the quality of the generated waveforms from the experimental models in a subjective evaluation test. 
In a single evaluation round, an evaluator listened to one speech waveform file on one screen, rated the speech quality on a 1-to-5 MOS scale,
and repeated the process for multiple screens.
The waveforms in one evaluation round were for the same text and were played in a random order.
Each evaluator could replay the waveform file during the evaluation. 
All the waveforms were converted to 16-bit PCM format in advance.

\begin{figure}[!t]
\centering
{\includegraphics[width=\columnwidth]{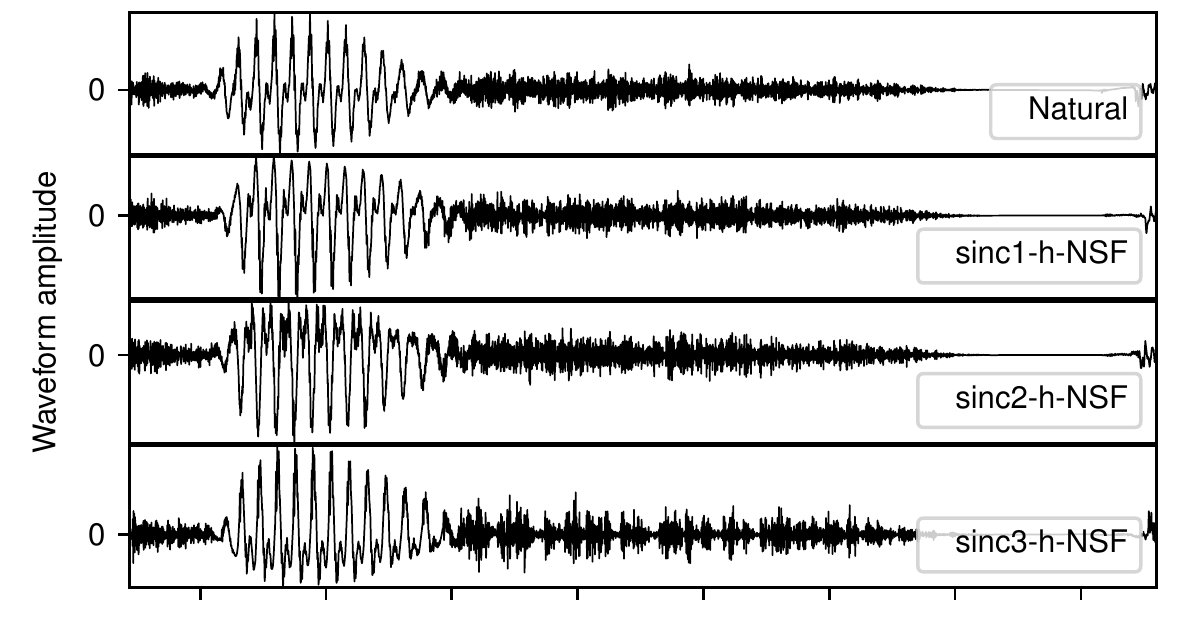}
}
\vspace{-6mm}
\caption{Natural and generated waveforms from models given natural acoustic features.}
\label{fig:waveform}
\end{figure}

Around 150 evaluators participated in the test, and 1604 sets of MOS scores were obtained.
The results plotted in Figure~\ref{fig:mos} demonstrate that \texttt{sinc1-h-NSF}, \texttt{base-h-NSF}, and \texttt{WaveNet} performed equally well. 
In contrast, \texttt{sinc2-h-NSF} and \texttt{sinc3-h-NSF} lagged behind. 
The reason for \texttt{sinc2-h-NSF}'s poor performance is the `under-estimated' MVF in voiced regions, as Figure~\ref{fig:mvf_ana} shows.
As a result, some voiced sounds generated by \texttt{sinc2-h-NSF} were over-aperiodic. For example, as Figure~\ref{fig:spec} plots, 
the voiced sound had a weak harmonic structure only around 4 kHz. 
 \texttt{sinc3-h-NSF} generated $\bs{f}_{1:T}^{(c)}=1$ for all utterances, and the waveforms generated from  \texttt{sinc3-h-NSF} lacked aperiodicity, which can be observed in Figure~\ref{fig:spec}. Furthermore, unvoiced sounds such as [s] were less aperiodic (see Figure~\ref{fig:waveform}) and sounded like a pulse train.

\begin{table}[t!]
\small
\caption{Number of network parameters and average number of waveform samples generated in 1-sec time on single Nvidia P100 GPU card.
\texttt{sinc*-h-NSF} had similar performance.}
\vspace{-5mm}
\begin{center}
\begin{tabular}{cccc}
\hline\hline
 & No. of model & \multicolumn{2}{c}{Generation speed} \\
\cline{3-4}
Model   & parameters & memory-save & normal mode \\
\hline
 \texttt{WaveNet} &  $2.96e+6$ & - & 0.19 k  \\
 \texttt{base-h-NSF} & $1.20e+6$ & 71 k & 335 k  \\
 \texttt{sinc1-h-NSF} & $1.20e+6$  & 70 k & 335 k \\
\hline\hline
\end{tabular}
\end{center}
\vspace{-6mm}
\label{tab:speed}
\label{tab:size}
\end{table}

Finally, Table~\ref{tab:speed} shows the number of parameters and the generation speed.
\texttt{WaveNet} was slow because of the AR generation process.
However, the NSF models were much faster because they produced the waveform in one shot.
In the memory-save mode, in which the NSF-models reduce GPU memory consumption by releasing and allocating memory layer by layer, 
the generation speed decreased because of the time for memory operation. However, they still surpassed \texttt{WaveNet}.

\section{Conclusion}
\label{sec:conclude}
We proposed a new h-NSF model with trainable MVF.
Compared with the baseline h-NSF model using pre-defined FIR filters to merge the harmonic and noise waveform components, the new h-NSF model predicts a time-variant MVF from the input acoustic features to adjust the frequency response of the FIR filters. 
We compared different strategies to predict the MVF in the experiments and found that the U/V information can be useful as prior knowledge. Specifically, we could predict a residual signal from the input acoustic features and add it to the U/V signal, which was more stable than other strategies such as directly predicting the MVF from scratch. Experiments demonstrated that the proposed trainable h-NSF can generate high-quality waveforms as good as the waveforms generated by WaveNet. Furthermore, the waveform generation speed of the proposed model was comparable to other NSF models and was much faster than that of WaveNet.

\noindent
\textbf{Acknowledgement:}
This work was partially supported by a JST CREST Grant (JPMJCR18A6, VoicePersonae project), Japan, and MEXT KAKENHI Grants (16H06302, 17H04687, 18H04120, 18H04112, 18KT0051), Japan. The experiments partially were conducted using TSUBAME 3.0 supercomputer of Tokyo Institute of Technology.

\bibliographystyle{IEEEtran}

\bibliography{../../BIB}

\end{document}